\documentclass{article}      
\usepackage{amsfonts,amssymb}
\newcount\commentcount \commentcount=0 
 \long\def\comment#1{\ifnum\commentcount=1 #1\fi}
 
\newtoks\reportnoregister \newtoks\eprintnoregister
\newcommand{\reportnumber}[1]{\reportnoregister={#1}}
\newcommand{\eprintnumber}[1]{\eprintnoregister={#1}}

\reportnumber{\mbox{}} 
\eprintnumber{\mbox{}} 

\newcommand{\reportid}{
   \begin{minipage}{17cm}\vspace{-3.2cm}
     \begin{flushright}
      {\normalsize \the\reportnoregister \\[-.2cm]
	    \eprintstyle{\the\eprintnoregister}}\vspace{3.2cm}
     \end{flushright}
   \end{minipage}\hspace{-17cm} }

\catcode`@=11   
\def\title#1{\gdef\@title{\reportid#1}}
\catcode`@=12   

\newcommand{\eprintstyle}[1]{\textsf{#1}} 

\newcommand{\journalfont}{\rm}  
\newcommand{\jou}[1]{{\journalfont #1\ }}
\newcommand{\joudef}[2]{\newcommand #1{\jou{\ignorespaces #2}}}

\joudef{\aaa}    { Astron.\ Astrophys.}
\joudef{\aip}    { Adv.\ Phys.}
\joudef{\adm}    { Adv.\ Math.}
\joudef{\am}     { Ann.\ Math.}
\joudef{\apb}    { Ann.\ Phys.\ (Berlin)}
\joudef{\apny}   { Ann.\ Phys.\ (N.Y.)}
\joudef{\apj}    { Astrophys.\ J.}
\joudef{\apjs}   { Astrophys.\ J.\ Suppl.}
\joudef{\cjp}    { Can.\ J.\ Phys.}
\joudef{\cmp}    { Commun.\ Math.\ Phys.}
\joudef{\cqg}    { Class.\ Quantum Grav.}
\joudef{\faa}    { Funct.\ Anal.\ Appl.}
\joudef{\grg}    { Gen.\ Rel.\ Grav.}
\joudef{\ijmpd}  { Int.\ J.\ Mod.\ Phys.\ D}
\joudef{\ijtp}   { Int.\ J.\ Theor.\ Phys.}
\joudef{\invm}   { Invent.\ Math.}
\joudef{\jm}     { J.\ Math.}
\joudef{\jmp}    { J.\ Math.\ Phys.}
\joudef{\jpa}    { J.\ Phys.\ A}
\joudef{\mnras}  { Mon.\ Not.\ R.\ Ast.\ Soc.}
\joudef{\mpla}   { Mod.\ Phys.\ Lett.\ A} 
\joudef{\nature} { Nature}
\joudef{\nc}     { Nuovo Cim.}
\joudef{\npb}    { Nuc.\ Phys.\ B}
\joudef{\ph}     { Physica}
\joudef{\pla}    { Phys.\ Lett. A}
\joudef{\plb}    { Phys.\ Lett. B}
\joudef{\pr}     { Phys.\ Rev.}
\joudef{\prd}    { Phys.\ Rev.\ D}
\joudef{\prep}   { Phys.\ Rep.}
\joudef{\prl}    { Phys.\ Rev.\ Lett.}
\joudef{\prsla}  { Proc.\ Roy.\ Soc.\ Lond.\ A}
\joudef{\ptp}    { Prog.\ Theor.\ Phys.}
\joudef{\ptps}   { Prog.\ Theor.\ Phys.\ Suppl.}
\joudef\rmp      { Rev.\ Mod.\ Phys.}
\joudef\spj      { Sov.\ Phys.\ JETP}

\newcommand{\mechgen}{03.20.+i}       
\newcommand{\symmcons}{11.30.-j}      
\catcode`@=11

\newcommand\eqalign[1]{\null\,\vcenter{\openup\jot\m@th
  \ialign{\strut\hfil$\displaystyle{##}$&$\displaystyle{{}##}$\hfil
      \crcr#1\crcr}}\,}
\newcommand\meqalign[1]{\null\,\vcenter{\openup\jot\m@th
  \ialign{\strut\hfil$\displaystyle{##}$&&$\displaystyle{{}##}$\hfil
      \crcr#1\crcr}}\,}
\def\ps@reportnumber{%
    \let\@oddfoot\@empty\let\@evenfoot\@empty
    \def\@oddhead{\hfil\rightmark}}
	
\catcode`@=12   

\newdimen\arrayruleHwidth
\makeatletter
\newcommand\Hline{\noalign{\ifnum0=`}\fi\hrule \@height \arrayruleHwidth
  \futurelet \@tempa\@xhline}
\makeatother

\newcommand\thickbaselines{\baselineskip=20pt\lineskip=3pt\lineskiplimit=3pt}

\catcode`@=11

\renewcommand\matrix[1]{\null\,\vcenter{\thickbaselines\m@th
    \ialign{\hfil$##$\hfil&&\quad\hfil$##$\hfil\crcr
      \mathstrut\crcr\noalign{\kern-\baselineskip}
      #1\crcr\mathstrut\crcr\noalign{\kern-\baselineskip}}}\,} 
\catcode`@=12   

 \newcommand\Hscr{{\cal H}}
 
\newcommand\Kscr{{\cal K}} 
 \newcommand\Nscr{{\cal N}}

\newcommand\undersim[1]{\mathop{\vtop{\ialign{##\crcr
     $\hfil\displaystyle{#1}\hfil$\crcr\noalign
     {\kern1pt\nointerlineskip}\hbox{$\hfil\sim\hfil$}\crcr
     \noalign{\kern1pt}}}}}

\begin{document}             
\bibliographystyle{prsty}     

\reportnumber{USITP 2001-8}
\eprintnumber{solv-int/0110021}

\title{A unified treatment of quartic invariants at fixed and  
arbitrary energy.} 
\author{ 
Max Karlovini\thanks{e-mail: max@physto.se} \\ 
Department of Physics -- Stockholm University \\ 
Giuseppe Pucacco\thanks{e-mail: pucacco@roma2.infn.it} \\ 
Dipartimento di Fisica -- Universit\`a di Roma ``Tor Vergata" \\ 
Kjell Rosquist\thanks{e-mail: kr@physto.se} \   
and 
Lars Samuelsson\thanks{e-mail: larsam@physto.se} \\ 
Department of Physics -- Stockholm University} 
\date{}
\maketitle 
\vspace{2cm}
\begin{abstract} 
Two-dimensional Hamiltonian systems admitting second invariants which  
are quartic in the momenta are investigated using the Jacobi  
geometrization of the dynamics. This approach allows for a unified  
 treatment of invariants at both arbitrary and fixed energy. In the  
differential geometric picture, the quartic invariant corresponds to  
the existence of a fourth rank Killing tensor. Expressing the Jacobi metric  
in terms of a K\"ahler potential, the integrability condition for the existence of  
the Killing tensor at fixed energy is a non-linear equation involving  
the K\"ahler potential. At arbitrary energy, further conditions must  
be imposed which lead to an overdetermined system with isolated  
solutions. We obtain several new integrable and superintegrable  
systems in addition to all previously known examples.     
\end{abstract}
 
\vspace{2cm}
\centerline{\bigskip\noindent PACS numbers: \mechgen \quad \symmcons }
\clearpage 
 
 
\section{Introduction} 
 
The study of Hamiltonian systems by means of geometric techniques
provides fruitful clues about their integrability
\cite{kolokoltsov:inv, geom, benenti}.  In particular, geometrization
of the dynamics by using the Jacobi metric \cite{arnold:mechanics,
  lanczos:mechanics}, is a standard tool to turn a natural Hamiltonian
system into a geodesic flow over a suitable Riemannian manifold.
Therefore, it is natural to investigate integrability of
two--dimensional systems by looking for invariants corresponding to
Killing tensors of a conformal Riemannian geometry. These geometric
objects in fact directly produce invariants polynomial in the momenta:
the degree of this polynomial is the same as the rank of the Killing
tensor. Since the energy enters in the Jacobi metric as a parameter,
the conditions for the existence of a Killing tensor may happen to be
satisfied only at a fixed value of energy. In this case we speak of
integrability at fixed energy in distinction to the usual notion of
integrability which is understood to be valid at arbitrary energy.
 
In \cite{geom}, quadratic invariants at arbitrary and fixed energy
(respectively, {\em strongly} and {\em weakly} conserved phase-space
functions) for 2-dimensional Hamiltonian systems were treated in a
unified way. The integrability condition for quadratic invariants
involves an arbitrary analytic function $S(z)$. For invariants at
arbitrary energy, the function $S(z)$ is a second degree polynomial
with real second derivative and the integrability condition then
reduces to the classical Darboux's condition for quadratic invariants
at arbitrary energy \cite{darboux:inv}. Thereafter, the possibility of
searching for linear and quadratic invariants at fixed energy was
addressed and some examples of system admitting a second quadratic
invariant at zero energy were provided.
 
Weak invariants, also called {\em configurational invariants}, have
been discussed by Hall \cite{hall:inv} and by Sarlet, Leach and
Cantrijn \cite{SLC85a}. Hietarinta \cite{hiet}, in his account of the
direct methods for the search of the second invariant, also provides a
review of all the known systems admitting one or more configurational
invariants.  Generalizing the approach of paper \cite{geom}, Karlovini
and Rosquist \cite{max} have discussed the existence of invariants
{\em cubic in the momenta} at both fixed and arbitrary energy. Beside
giving a list of all known systems admitting a cubic strong invariant,
they also find a superintegrable system admitting a cubic invariant
related to an energy dependent linear invariant. In the present paper
we are going to discuss the case of the quartic invariant associated
with the existence of a fourth-rank Killing tensor. Fourth-rank
Killing tensors have previously been studied by one of the authors
(see \cite{lars}) in the Lorentzian case.
 
In analogy to the third--rank case, it turns out that it is natural to
introduce a K\"ahler potential for the Jacobi metric. Expressing the
conditions imposed on the geometry by Killing tensor equations in
terms of this K\"ahler potential, there remains a system of nonlinear
partial differential equations. Imposing the condition of strong
integrability, it turns out that in general this system becomes
overdetermined, so that, like for cubic invariants, only isolated
mechanical systems exist with a second invariant quartic in the
momenta. Due to the complexity of the system of equations, it is
impossible to get a fully general solution. However, with our approach
we are able to find some new integrable and superintegrable cases. In
particular, one may wonder if the knowledge of a weak invariant can
provide information about the global dynamical behaviour of the
system. We introduce a method which, starting from a family of weakly
integrable systems, leads to finding a higher-order invariant for
isolated members of the family which are therefore strongly
integrable. The second invariant depends on the energy as a parameter.
 
The plan of the paper is as follows: in section 2 we recall the
necessary and sufficient conditions for the existence of a Killing
tensor in the Jacobi geometry and their application in problems of
analytical mechanics; in section 3 we examine the particular case of
fourth--rank Killing tensors; in section 4 we analize the conditions
for strong integrability with a quartic invariant; in section 5 we
present the method by which, starting from a family of weakly
integrable systems, isolated examples of strongly integrable systems
with quartic second invariant can be found; in section 6 we give
tables of integrable systems admitting a quartic invariant which
include new integrable and superintegrable systems in addition to all
known cases; in section 7 we give our conclusions.

 
\section{Jacobi geometry and Killing tensors} 
 
 
\subsection{Geometric representation of the dynamics} 
 
We are interested in the classical two-dimensional systems with
Hamiltonian function
\begin{equation}\label{H} 
\Hscr = {1 \over 2} (p_x^2 + p_y^2) + V (x, y). 
\end{equation} 
The approach followed in studying the integrability properties of
these systems is based on the Jacobi geometrization procedure, which
associates to the Hamiltonian flow produced by (\ref{H}), a geodesic
flow on a Riemannian manifold endowed with a positive definite metric
\begin{equation} 
ds_J^2 = g_{\alpha \beta} d q^{\alpha} d q^{\beta}, \quad \alpha, \beta = 1,2. 
\end{equation} 
It can be shown (see, e.g. \cite{arnold:mechanics}), that the Jacobi
metric
\begin{equation} 
g_{\alpha \beta} = 2(E - V) h_{\alpha \beta} ,  
\end{equation} 
where $h_{\alpha \beta}$ is the metric of the flat space, generates a
geodesic flow corresponding to the natural mechanical system with
potential $V = V(q)$ at energy $E$. In fact, introducing the geodesic,
or ``Jacobi", Hamiltonian
\begin{equation} 
{\Hscr}_J = {1 \over 2} {1 \over {2(E - V)}}  
              h^{\alpha \beta} p_{\alpha} p_{\beta} \equiv {1 \over  
2},  
\end{equation} 
the geodesic equations 
\begin{equation}\eqalign 
{{{d q^{\alpha}} \over {ds_J}} &=  
 {{\partial {\Hscr}_J} \over {\partial p_{\alpha}}} = 
 {1 \over {2(E - V)}} h^{\alpha \beta} p_{\beta}, \cr 
 {{d p_{\alpha}} \over {ds_J}} &= -  
 {{\partial {\Hscr}_J} \over {\partial q^{\alpha}}} = - 
 {1 \over {2(E - V)}} {{\partial V} \over {\partial q^{\alpha}}}, \cr} 
\end{equation} 
are equivalent to the canonical equations of motion given by the  
Hamiltonian 
(\ref{H}),  
\begin{equation}\eqalign 
{{{d q^{\alpha}} \over {dt}} &=  
 {{\partial {\Hscr}} \over {\partial p_{\alpha}}}, \cr 
 {{d p_{\alpha}} \over {dt}} &= -  
 {{\partial {\Hscr}} \over {\partial q^{\alpha}}}, \cr} 
\end{equation} 
if the 
natural identification of canonical coordinates 
\begin{equation} 
q^1 = x, \;\; q^2 = y, \;\; p_1 = p_x, \;\; p_2 = p_y  
\end{equation} 
is made. The phase-space trajectories are parametrized by the affine
parameter $s_J$ that is related to the standard time variable by
\begin{equation} 
ds_J = 2[E - V(q(t))] dt. 
\end{equation} 
The Jacobi geometry corresponding to Hamiltonian (\ref{H}) is then 
\begin{equation}\label{G} 
ds^2 = 2G(dx^2+dy^2)=2Gdzd\bar z, \;\; G=E-V, 
\end{equation} 
where we have introduced null variables $z=x+iy$ and $\bar z = x-iy$.
 
All tensor calculations will be done in the standard null frame
defined as
\begin{equation} 
\Omega^0 = \sqrt{G} dz, \;\; 
\Omega^{\bar 0} = \sqrt{G} d\bar z, 
\end{equation} 
in which the metric takes the simplest possible form 
\begin{equation} 
ds^2 = 2 d\Omega^0 d\Omega^{\bar 0}. 
\end{equation} 
We use the convention that tensor indices in this frame take the
values $0$ and $\bar 0$, while in any coordinate frame the values will
be the names of the coordinates (e.g. $z$ and $\bar z$).
 
 
\subsection{Invariants polynomial in the momenta and Killing tensors} 
 
One of the standard tools of the geometric approach is the
investigation of integrability by looking for invariants generated by
{\it Killing tensors}. Let us see the simple case of second-order
Killing tensor equations
\begin{equation}\label{KE2} 
K_{(\alpha \beta ; \gamma)} = 0. 
\end{equation} 
The existence of a second--rank Killing tensor, that is a symmetric
tensor which satisfies eq.(\ref{KE2}), implies the existence of a
conserved quadratic function
\begin{equation}\label{KTQ} 
I_J = K^{\alpha \beta} p_{\alpha} p_{\beta} 
\end{equation} 
commuting with the Jacobi Hamiltonian. In fact, it is easy to check
that the vanishing of the Poisson bracket of this function with the
Hamiltonian implies eq.(\ref{KE2}). In analogy with the theory of
second order invariants, the Jacobi geometry approach leads, as it
stands, to the determination of higher-order invariants. Generalizing
eq.(\ref{KE2}), an {\it m'th rank} Killing tensor is a symmetric
tensor satisfying equation
\begin{equation}\label{KEM} 
K_{(\mu_1 \mu_2 ... \mu_m ; \mu_{m + 1})} = 0. 
\end{equation} 
It gives rise to the {\it m'th order} (in the momenta) invariant 
\begin{equation}\label{KTM} 
I_J = K^{\mu_1 \mu_2 ... \mu_m} p_{\mu_1} ... p_{\mu_m}. 
\end{equation} 
Two remarks are in order here: 
 
1) The function (\ref{KTM}) is a {\it weak invariant}, in the sense
that, in general, it is a conserved quantity in the dynamics fixed by
the given value of the energy appearing in the definition of the
Jacobi geometry (\ref{G}).  To grant it the full title of strong
invariant, it must satisfy the additional requirement of satisfying
the Killing tensor equations at arbitrary energy.
 
2) The function (\ref{KTM}), let us call it the {\it Jacobi
  invariant}, is a homogeneous polynomial of degree equal to the rank
of the corresponding Killing tensor. To transform it into the ordinary
invariant in the physical time gauge there is a straightforward recipe
consisting in replacing the parameter $E$ appearing in $I_J$ with the
corresponding Hamiltonian function (\ref{H}). As a consequence, the
physical invariant
\begin{equation}\label{prescription} 
I = I_J \big\vert_{E \rightarrow {\cal H}} 
\end{equation} 
becomes a polynomial which is either even or odd in the momenta. 
 
Karlovini and Rosquist (2000, \cite{max}) have discussed the existence
of third-rank Killing tensors, giving a list of all known integrable
or superintegrable systems admitting a cubic strong invariant. Here we
are going to discuss the case of the quartic invariant associated with
the existence of a fourth-rank Killing tensor. Moreover, we want to
exploit the results concerning the existence of higher-order
invariants to generate new solutions starting from the assumption of
the existence of a second-rank Killing tensor.
 
 
\section{Quartic invariants corresponding to fourth rank Killing  
tensors} 
 
In this section we derive the necessary and sufficient integrability
condition for the Jacobi metric to admit a fourth-rank Killing tensor
at a fixed value of the energy parameter $E$. In the following section
we proceed by finding the conditions that ensure that the Killing
tensor equations be satisfied at arbitrary values of the energy, with
the restriction to the case in which the energy dependence of the
Killing tensor is such that the corresponding invariant of the
physical Hamiltonian is quartic in the momenta. As usual in our
approach (see \cite{lars}), we decompose the fourth-rank Killing
tensor in the form:
 
\begin{equation}\label{KT1}  
K_{\alpha\beta\gamma\delta}=P_{\alpha\beta\gamma\delta} 
                           +P_{(\alpha\beta}g_{\gamma\delta)} 
                 +{3\over8}Kg_{(\alpha\beta}g_{\gamma\delta)} 
\end{equation} 
or equivalently 
\begin{equation}\label{KT2} 
K_{\alpha\beta\gamma\delta}=P_{\alpha\beta\gamma\delta} 
                           +K_{(\alpha\beta}g_{\gamma\delta)} 
                 -{1\over8}Kg_{(\alpha\beta}g_{\gamma\delta)} 
\end{equation} 
where $K_{\alpha\beta}=K^{\gamma}{}_{\alpha\beta\gamma}$ and $ K =
K^{\gamma}{}_{\gamma}$. $P_{\alpha\beta\gamma\delta}$ is the
trace-free (conformal) part,
$P_{\alpha\beta}=P^{\gamma}{}_{\alpha\beta\gamma}$ its trace and $K$
the full trace of the Killing tensor. In this way, the Killing tensor
equations are split into the trace-free ``components" and the equation
for the trace
\begin{equation}\label{KTE1} 
          P_{(\alpha\beta\gamma\delta;\mu)} 
-{1\over2}P^{\nu}{}_{(\alpha\beta\gamma;\nu}g_{\delta\mu)}=0,  
\end{equation} 
 
\begin{equation}\label{KTE2} 
          P_{(\alpha\beta;\gamma)} 
-{1\over2}P^{\nu}{}_{(\alpha;\nu}g_{\gamma\delta)} 
+{1\over2}P^{\nu}{}_{\alpha\beta\gamma;\nu}=0,  
\end{equation} 
 
\begin{equation}\label{KTE3} 
K_{,\alpha} = - {4 \over 3} P^{\beta}{}_{\alpha;\beta}.  
\end{equation} 
It is advantageous to employ also the coordinate frame components of
the conformal part and of its trace when parametrizing the five
independent components of $K_{\alpha\beta\gamma\delta}$. Thus, the
components of the Killing tensor can be written as
\begin{equation} 
\eqalign{K_{0000} & = G^2 P^{\bar{z}\bar{z}\bar{z}\bar{z}} =  
G^2\overline{S} \cr 
         K_{000 \bar 0} & = {1\over2}P_{00} = {1\over2} G P^{\bar{z}\bar{z}} =  
           {1\over2}G\overline{R(z,\bar{z})} \cr 
         K_{00 \bar 0 \bar 0} & = {1\over4}K \cr 
         K_{0 \bar 0 \bar 0 \bar 0} & = {1\over2}P_{\bar 0 \bar 0} = 
                                  {1\over2} G P^{zz} =  
                                  {1\over2}G{R(z,\bar{z})}\cr      
         K_{\bar 0 \bar 0 \bar 0 \bar 0} & = G^2 P^{zzzz} = G^2 S,\cr} 
\end{equation} 
where, as usual, $S = S(z)$ is a holomorphic function, so that
$\overline{S} = \overline{S}(\bar{z})$ comes from (\ref{KTE1}), but
the function $ R = R(z,\bar{z}) $ is a generic function of the
arguments. With this parametrization, the equation for the trace
(\ref{KTE3}) has the integrability condition
 
\begin{equation}\label{intcon4} 
2(G_{,zz} R - G_{,\bar z \bar z} \bar R) + 
3(G_{,z} R_{,z} - G_{,\bar z} \bar R_{,\bar z}) + 
G(R_{,zz} - \bar R_{,\bar z \bar z}) =   0, 
\end{equation}  
which is analogous to that obtained for the second rank case (but now
$R$ is not holomorphic!).
 
We now proceed to solve the Killing tensor equations. Eq.(\ref{KTE2})
gives
\begin{equation}\label{EQR} 
S^{-3/4} R_{, \bar z} + 2 (S^{1/4} G)_{,z} = 0  
\end{equation} 
and its complex conjugate. In analogy with the approach followed in
\cite{geom, max}, we make a coordinate transformation to put the
conformal Killing tensor in the simplest ({\em standard}) form.  Using
a conformal transformation of the form
\begin{equation}\label{CC} 
\eqalign{w &= H(z) , \cr w &=  X + i Y , \cr 
      z &= x + i y , \cr}  
\end{equation} 
the transformation of the conformal tensor is such that
\begin{equation} 
\tilde{S}(w) := P^{wwww} = [H'(z)]^4 P^{zzzz} = [H'(z)]^4 S(z). 
\end{equation} 
If we make the standard choice 
\begin{equation} 
\tilde{S}(w) := P^{wwww} = 1, 
\end{equation} 
the conformal transformation is then provided by the function
\begin{equation}\label{transf} 
H'(z)=[S(z)]^{-1/4}.  
\end{equation} 
Eq.(\ref{EQR}) becomes 
\begin{equation}\label{EQRW} 
(S^{-1/2} R)_{, \bar w} + 2 \ {\tilde G}_{,w} = 0, 
\end{equation} 
with  
\begin{equation} 
{\tilde G} = {|H'(z)|}^{-2} G = \sqrt{|S|} \ G. 
\end{equation}  
Moreover, under the conformal transformation generated by
eq.(\ref{transf}), the function $R$ transforms as
\begin{equation}\label{EQRR} 
\tilde R = P^{ww} = [H'(z)]^2 P^{zz} = S^{-1/2} R. 
\end{equation}  
Inserting $\tilde R$ into eq.(\ref{EQRW}), we have the set 
\begin{equation}\label{EQRS} 
\eqalign{ 
           \tilde R_{, \bar w} + 2 \ {\tilde G}_{,w} &= 0, \cr 
{\overline{\tilde R}}_{,w}     + 2 \ {\tilde G}_{,\bar w} &= 0. \cr} 
\end{equation} 
The solution of this system can be found in terms of a 
real K\"ahler potential $\Kscr(w, \bar{w})$: 
\begin{equation}\label{Kahler} 
\eqalign 
{\tilde G &= {\Kscr}_{, w \bar{w}}, \cr 
 \tilde R &= - 2 \ {\Kscr}_{, w w}. \cr} 
\end{equation} 
whereas the integrability condition (\ref{intcon4}) determines the
following equation for the K\"ahler potential:
\begin{equation}\label{intcon} 
\Im \bigl\{ ({\Kscr}_{, w w w} {\Kscr}_{, w \bar{w}} + 2  
 {\Kscr}_{, w w \bar{w}} {\Kscr}_{, w w} )_{, w} \bigr\} = 0. 
\end{equation} 
This is the necessary and sufficient condition for the existence of a
fourth rank Killing tensor. In analogy to the third rank case (see
\cite{max}) and in contrast to the first and second rank case
\cite{geom}, the condition is highly nonlinear. The same condition has
been already found in \cite{hall:inv} and treated in \cite{hiet}, even
if not in the context of the present geometric approach. Note that in
\cite{hiet}, eq.(7.5.15), there is a misprint for a factor 2 missing.
 
In the standardized coordinate frame the second invariant can be
written in the form
\begin{equation}\label{qwi} 
I_J = 2 \ \Re \{p_w^4 + 2 \ \tilde R \ {\Hscr}_J \ p_w^2 \} 
      + {3 \over 2} {\Hscr}_J^2 K,  
\end{equation} 
whereas in the original null coordinate frame the second invariant is
\begin{equation}\label{qzi} 
I_J = 2 \ \Re \{S \ p_z^4 + 2 \ R \ {\Hscr}_J \ p_z^2 \} 
      + {3 \over 2} {\Hscr}_J^2 K,  
\end{equation} 
where 
\begin{equation} 
{\Hscr}_J = {1 \over G} p_{z} p_{\bar z}  
          = {1 \over {\tilde G}} p_{w} p_{\bar w} 
          = p_0 p_{\bar 0} 
            \equiv {1 \over 2}, 
\end{equation} 
is the Jacobi Hamiltonian expressed in the three different reference
frames. The trace function $K$ is found by integrating the system
\begin{equation}\label{trace} 
\eqalign{ 
{3 \over 4} K_{, \bar{w}} + \tilde R_{,w} \tilde G + 2 \tilde R  
\tilde G_{,w} 
&= 0, \cr  
{3 \over 4} K_{, w} + {\bar {\tilde R}}_{,\bar w} \tilde G + 2 {\bar  
{\tilde R}}  
\tilde G_{,\bar w} &= 0. \cr} 
\end{equation} 
 
 
\section{Arbitrary energy invariants} 
 
Eq.(\ref{intcon}) gives the the necessary and sufficient condition for
a two-dimensional Riemannian geometry to admit a fourth-rank Killing
tensor. If we interpret the geometry according to eq.(\ref{G}) as that
providing a natural mechanical system, we can derive the additional
conditions that make the Killing tensor equations satisfied for every
value of the energy $E$. We use the following form for the K\"ahler
potential,
\begin{equation}\label{Kansatz} 
\Kscr = E [ z \bar z + 2 \ \Re \{ \Lambda (z) \} ] - \Psi, 
\end{equation} 
where $\Lambda$ is a holomorphic function independent of $E$ and the
real {\em pre--potential} $\Psi$ is such that
\begin{equation} 
\Psi_{,z\bar z} = V. 
\end{equation} 
 
In fact, no loss of generality is implied by letting the energy
dependence of the K\"ahler potential be prescribed by
eq.(\ref{Kansatz}) and by taking the function $S_4(z)$ to be energy
independent, given that we only take interest in the cases for which
the physical invariant $I$ is a quartic polynomial in the momenta.
Since the analogous statement was made without proving it for the
cubic case \cite{max}, we give a proof here. We begin by noting that
the physical invariant $I$, obtained from the Jacobi invariant $I_J$
of eq.(\ref{qzi}) according to prescription (\ref{prescription}), can
be written as
\begin{equation}\label{inv} 
  I = 2\Re \left\{S p_z^4 + B p_z^3 p_{\bar{z}}\right\} + C (p_z p_{\bar{z}})^2, 
\end{equation} 
where 
\begin{equation}\label{BandC} 
\eqalign{ 
  B &= 2RG^{-1}, \cr 
  C &= \frac32 K G^{-2}.\cr} 
\end{equation} 
Since the energy parameter $E$ is now assumed to have been replaced by
the physical Hamiltonian $\Hscr$, the functions $S$, $B$ and $C$ will
be dependent on the momenta $p_z$ and $p_{\bar{z}}$, but clearly only
through the combination $p_z p_{\bar{z}}$ that appears in $\Hscr$.
This implies that all five terms on the right hand side of
eq.(\ref{inv}) have to be quartic polynomials in the momenta since $I$
is quartic by definition. This can be clearly seen by viewing $I$ as a
function of the two independent momenta functions $p_z p_{\bar{z}}$
and $p_z/p_{\bar{z}}$, rather than the momenta $p_z$ and $p_{\bar{z}}$
themselves. Indeed, $I$ can be recast into the form
\begin{equation} 
  I = (p_z p_{\bar{z}})^2\sum_{k=-2}^2 Q_k(p_z/p_{\bar{z}})^k, 
\end{equation} 
where  
\begin{equation} 
\eqalign{ 
  &Q_{-2} = \overline{{Q}_2} = S, \cr 
  &Q_{-1} = \overline{{Q}_1} = B, \cr 
  &Q_0 = C. \cr} 
\end{equation} 
Clearly, with the coefficients of $(p_z/p_{\bar{z}})^k$ only depending
on the momenta through $p_z p_{\bar{z}}$, there can be no cancellation
of possible non-quartic polynomial dependence in the individual terms.
Hence, we conclude that the functions $Sp_z^4$, $Bp_z^4p_{\bar{z}}$
and $C(p_zp_{\bar{z}})^2$, as well as the complex conjugates of the
first two, must be quartic momenta polynomials. It immediately follows
that $S$, $B$ and $C$ are polynomials in $(p_z p_{\bar{z}})^{-1}$,
with coefficients depending only on $z$ and $\bar{z}$, of degree zero,
one and two, respectively. This directly proves the part of statement
about $S$ having no energy dependence. Moreover, since $G = E - V$
turns into $T = 2p_zp_{\bar{z}}$ when $E$ is replaced by $\Hscr = T +
V$, it now follows from eqs.(\ref{BandC}) that the functions $R$ and
$K$ are restricted to be first respectively second degree polynomials
in $p_z p_{\bar{z}}$, which is the same as saying that they must be
the same type of polynomials in the energy parameter $E$, before
making the substitution $E \rightarrow \Hscr$. Using the second of
eqs.(\ref{Kahler}) and the conformal transformation formulae given by
eqs.(\ref{transf}) and (\ref{EQRR}), the function $R$ can be expressed
in terms of $S$ and $\mathcal{K}$ according to
\begin{equation}\label{SKgivesR} 
  R = -2(S\mathcal{K}_{,zz} + \frac14 S'\mathcal{K}_{,z}), 
\end{equation} 
while eqs.(\ref{trace}), which notably are form invariant under conformal 
transformations, give that the trace $K$ is given by integrating 
\begin{equation}\label{SKgivesK} 
  \frac34 K_{,\bar{z}} + R_{,z}G + 2RG_{,z} = 0 
\end{equation} 
and its complex conjugate. Now, there is clearly no loss in generality 
to write the K\"ahler potential $\Kscr$ as 
\begin{equation} 
\Kscr = E[z\bar{z} + 2\Re \{\Lambda(E,z) \}] - \Psi(z,\bar{z}), 
\end{equation} 
with $\Lambda$ being analytic in $z$ and $\Psi$ being energy 
independent. Moreover, since $S$ is energy independent and $R$ is a 
first degree polynomial in $E$, it follows from eq.(\ref{SKgivesR}) 
that we are able to write $\Lambda(E,z)$ as 
\begin{equation} 
  \Lambda(E,z) = \Lambda_0(z) + \Lambda_1(E,z),  
\end{equation} 
where $\Lambda_1(E,z)$ does not contribute to eq.(\ref{SKgivesR}). 
However, since eq.(\ref{SKgivesK}) shows that $\Lambda(z)$ enters 
into $K_{,\bar{z}}$ only through $R$, we might as well set 
$\Lambda_1(E,z)$ to zero, which proves the fact that we could have 
taken $\Lambda(E,z)$ to be energy independent from the outset. We may 
finally note that with both $R$ and $G$ being first degree polynomials 
in $E$, we find from eq.(\ref{SKgivesK}) that $K_{,\bar{z}}$ is a 
second degree polynomial in $E$, which indeed is a necessary and 
sufficient condition for $K$ to be a second degree polynomial as well, 
up to addition of an irrelevant integration constant which in 
principle can have an arbitrary energy dependence. This completes the  
proof that the form of the K\"ahler potential given by  
eq.(\ref{Kansatz}) is the most general one which is needed for full  
generality. 
 
We are now in the position to solve the integrability condition at
arbitrary energy. In the original $z$ coordinates, eq.(\ref{intcon})
takes the form
\begin{equation}\label{zintcon} 
\Im \Bigl\{ \bigl[ 
S \bigl({\Kscr}_{, zzz} {\Kscr}_{, z \bar{z}} + 2   
        {\Kscr}_{, zz\bar z} {\Kscr}_{, zz} \bigr) + {1 \over 2}  
S'\bigl({\Kscr}_{, zz \bar z} {\Kscr}_{, z} + {5 \over 2}  
{\Kscr}_{, zz} {\Kscr}_{, z \bar{z}} \bigr) + {1 \over 4}  
S'' {\Kscr}_{, z \bar{z}} {\Kscr}_{, z} \bigr]_{,z} \Bigr\} = 0. 
\end{equation} 
Substituting the ansatz (\ref{Kansatz}) into eq.(\ref{zintcon}), this
condition provides a second degree polynomial in $E$:
\begin{equation}\label{poli} 
A_2 E^2 + A_1 E + A_0 = 0. 
\end{equation} 
The coefficients $A_0, A_1, A_2$ must vanish separately if the
equation must hold for arbitrary energy. Therefore we obtain the three
equations:
\begin{equation}\label{A2} 
A_2 = \Im \Bigl\{ \bigl[ \Lambda''' S +  
             {5 \over 4} \Lambda''  S' +   
{1 \over 4} (\bar z +    \Lambda' ) S'' ]_{,z} \Bigr\} = 0,  
\end{equation} 
\begin{equation}\label{A1} 
\eqalign{ 
A_1 &= \Im \Bigl\{ \bigl[ 
S \bigl(\Psi_{, zzz} + 2 \Lambda''' \Psi_{, z \bar z} 
                     + 2 \Lambda''  \Psi_{, z z \bar z}\bigr) \cr  
&+ {1 \over 2} S'  
        \bigl( (\bar z + \Lambda') \Psi_{, zz \bar z} + {5 \over 2}   
(\Psi_{, zz} + \Lambda'' \Psi_{, z \bar z}) \bigr) \cr 
&+ {1 \over 4} 
S'' \bigl( \Psi_{,z} + (\bar z + \Lambda') \Psi_{, z \bar z}  
\bigr) \bigr]_{,z} \Bigr\} = 0, \cr} 
\end{equation} 
\begin{equation}\label{A0} 
A_0 = \Im \Bigl\{ \bigl[ 
S \bigl(\Psi_{, zzz} \Psi_{, z \bar{z}} + 2  
\Psi_{, zz \bar z} \Psi_{, z z} \bigr) + {1 \over 2} S'  
\bigl(\Psi_{, zz \bar z} \Psi_{, z} + {5 \over 2} \Psi_{, z \bar z}  
\Psi_{, z z} 
\bigr) + {1 \over 4} S''  
\Psi_{, z \bar z} \Psi_{, z} \bigr]_{,z} \Bigr\} = 0.  
\end{equation} 
It turns useful to express the system of eqs.(\ref{A2}--\ref{A0}) also
in the transformed coordinates since this simplifies the computations
when the $S$ function is of higher degree. Using the notation
\begin{equation}\label{ict} 
F(w) = H^{-1} (z(w)) 
\end{equation} 
for the inverse conformal transformation, the coefficients $A_0, A_1,
A_2$ become
\begin{equation}\label{B2} 
A_2= \Im \Bigl\{ \bigl[ F''' \bar F F' \bar F'  +  
                        2 (F'')^2 \bar F \bar F'  +  
                         \Lambda''' F' \bar F' +  
                         \Lambda''  F'' \bar F' \bigr]_{,w} \Bigr\},   
\end{equation} 
\begin{equation}\label{B1} 
A_1= \Im \Bigl\{ \bigl[ (F''' \bar F + \Lambda''') \Psi_{, w \bar w}  
      + 2 (F'' \bar F + \Lambda'') \Psi_{, w w \bar w} 
      + F' \bar F' \Psi_{, w w w} + 2 F'' \bar F' \Psi_{, w w}  
\bigr]_{,w} 
         \Bigr\},  
\end{equation} 
\begin{equation}\label{B0} 
A_0 = \Im \Bigl\{ \bigl[\Psi_{, w w w} \Psi_{, w \bar{w}} + 2  
                       \Psi_{, w w \bar{w}} \Psi_{, w w} \bigr]_{,w} 
          \Bigr\}. 
\end{equation} 
Applying the differential operator  
$$  
{{ \partial^2} \over {\partial z \partial \bar z}} 
$$ 
to eq.(\ref{A2}), we get the condition 
\begin{equation}\label{S4} 
\Im \{ S''''(z) \} = 0, 
\end{equation} 
so that the form allowed to the $S$ function to have integrability at
arbitrary energy is
\begin{equation} 
S(z) = a z^4 + \beta z^3 + \gamma z^2 + \delta z + \epsilon, \quad 
a \in \mathbb{R}, \quad \beta, \gamma, \delta, \epsilon \in \mathbb{C} 
\end{equation} 
In \cite{geom} it has been shown that, to get arbitrary energy
quadratic invariants in systems with two degrees of freedom, the
function $S(z)$ must satisfy the condition
\begin{equation}\label{S2} 
\Im \{ S_2''(z) \} = 0, 
\end{equation} 
that is it must be a second degree polynomial with real second
derivative. Together with the result (\ref{S4}) and the corresponding
one obtained in the third--rank case \cite{max},
\begin{equation}\label{S3} 
\Re \{ S_3'''(z) \} = 0, 
\end{equation} 
we can guess that, as a general rule, the analytic function $S(z)$
representing the conformal part of a Killing tensor of arbitrary rank
$m$ is required to satisfy the condition
\begin{equation} 
\eqalign{ 
\Re \Bigl\{ \bigl({d \over {dz}} \bigr)^m S(z) \Bigr\} &= 0  
\;\; (m \; {\rm odd}),\cr 
\Im \Bigl\{ \bigl({d \over {dz}} \bigr)^m S(z) \Bigr\} &= 0  
\;\; (m \; {\rm even}).\cr} 
\end{equation} 
As a simple illustration of the solution of the above set of  
eqs.(\ref{A2})--(\ref{A0}), let us take 
$$ 
\Lambda = 0, \; \; S = b, \; b \in \mathbb{R}. 
$$ 
Eq.(\ref{A2}) is automatically satisfied. Eq.(\ref{A1}) gives then the equation 
$$ 
\Psi_{,xxxy} - \Psi_{,xyyy} = 0, 
$$ 
whose solution is 
$$ 
\Psi_{,xy} = F_1 (x + y) + F_2 (x - y). 
$$ 
Therefore 
$$ 
\Psi = f_1 (x) + f_2 (y) + f_3 (x - y) + f_4 (x + y). 
$$ 
The potential can then be written as 
$$ 
V = v_1 (x) + v_2 (y) + v_3 (x - y) + v_4 (x + y), 
$$ 
where $ v_i = f_i'' / 4, \ i=1,2$ and $ v_i = f_i'' / 2, \ i=3,4$. Eq.(\ref{A0}) now gives 
$$ 
\eqalign{ 
&(v_1'' v_4 + 2 v_4'' v_1) + 3 v_1' v_4' - 
 (v_1'' v_3 + 2 v_3'' v_1) - 3 v_1' v_3' -\cr 
&(v_2'' v_4 + 2 v_4'' v_2) - 3 v_2' v_4' + 
 (v_2'' v_3 + 2 v_3'' v_2) - 3 v_2' v_3' = 0.\cr} 
$$
This equation coincides with that reported in the review of
Hietarinta \cite{hiet}.  Its most relevant solutions, together with
the expressions of the invariant, are listed in table 2 below.
 
In more general cases eq.(\ref{A1}) is not as easily solved. The
standard approach is to restrict the attention to special cases where
$S(z)$ is either a homogeneous polynomial or a non-homogeneous one
which is simply a power of the corresponding function in lower rank
cases. Even with such restrictions, many solutions can be obtained.
Moreover, in all the solutions found it has turned out that $\Lambda
(z) = \lambda z^2$, with $\lambda$ a complex constant. In this way the
coefficient
$$  
z \bar z + 2 \ \Re \{ \Lambda (z) \}  
$$
of $E$ in the K\"ahler potential (\ref{Kansatz}) is always a
Hermitian form in $z$.  In the tables included in Section 6 are listed
all the solutions found.
 
 
\section{Strong invariants generated by lower-order weak invariants} 
 
In the present section we want to introduce an alternative technique
to identify classes of strongly integrable systems. This approach
works only in a restricted subclass of systems, but is useful to get
insight into the structure of general solutions. The key point of the
idea is based on the construction of a generic weak integrable system
corresponding to the existence of a second-rank Killing tensor.
Introducing the corresponding simplified K\"ahler potential in the
general set of equations, we try to isolate single systems that
satisfy them. The reader more interested in the general approach may
wish to skip directly to section 6 with the results.
 
 
\subsection{Quadratic invariants} 
 
The technique devised and applied in paper \cite{geom} amounts, in
short synthesis, in a conformal transformation of the form (\ref{CC}).
In the second rank case, it turns out that it must be generated by an
arbitrary holomorphic function $S_2(z)$ via the relation
\begin{equation}\label{dct} 
H(z) = \int {dz \over {\sqrt{S_2(z)}}} . 
\end{equation} 
In the new coordinates $X,Y$, the Jacobi potential $ G = E - V $ turns
out to be of the form
\begin{equation}\label{G(X,Y)} 
G(X,Y) = {{A(X) + B(Y)} \over |S_2(X,Y)|} , 
\end{equation} 
where $A$ and $B$ are arbitrary functions of their arguments, which
can be thought of as the {\it separation coordinates}. The conformal
factor appearing in (\ref{G(X,Y)}) can be written as
\begin{equation}\label{confact} 
|S_2(X,Y)| = \sqrt{ S_2(w) \bar S_2 (\bar w)} = F' (w) \bar F' (\bar  
w). 
\end{equation}  
Moreover, we can write the transformation of the momenta in the form
\begin{equation}\eqalign{ 
      p_X &=   R \ p_x + Q \ p_y , \cr 
      p_Y &= - Q \ p_x + R \ p_y , \cr  
}\end{equation} 
where  
\begin{equation}\eqalign{ R = \Re \{ F^{\prime} \} , \cr  
             Q = \Im \{ F^{\prime} \} . \cr} 
\end{equation} 
We can therefore write the expression of the second invariant in the
Jacobi gauge, as
\begin{equation}\label{JSI} 
I_S (p_X, p_Y, X, Y) = {1 \over 2} (p_X^2 - p_Y^2) + B(Y) - A(X). 
\end{equation} 
As remarked in the introduction, since to each Jacobi potential $G$
pertains a specific dynamical system, we have that the found invariant
is, in general, only a {\it weak} invariant for the standard
Hamiltonian, in the sense that it provides a conserved quantity only
at a given value of the energy (perhaps at {\it zero} energy). In
fact, the Poisson bracket of the function (\ref{JSI}) with the
separated Hamiltonian
\begin{equation}\label{SH} 
{\Hscr}_S = {{{1 \over 2} (p_X^2 + p_Y^2) - A(X) - B(Y)}  
             \over |S_2|}, 
\end{equation} 
is 
\begin{equation}\label{poisson1} 
\{I_S, {\Hscr}_S \} = { E \over |S_2|} (p_X \ |S_2|_{,X} - p_Y \  
|S_2|_{,Y}).  
\end{equation}

 
\subsection{Quartic strong invariants generated by quadratic weak  
invariants} 
 
In section 4, the system of equations (\ref{A2})--(\ref{A0}) was
treated and many non trivial solutions were found. A general solution
is however very difficult to find due to the complicated structure of
the full system. Here we want instead to exploit the above results to
generate new solutions simply starting from the assumption of the
existence of a second-rank Killing tensor.
 
As a foreword to this approach, let us make the following remark. Let
us suppose to have made the choice of the $S_2$ function generating
the conformal transformation which gives the second invariant
(\ref{JSI}) satisfying equation (\ref{poisson1}). Denoting by
${\Hscr}_0$ the numerator appearing in the expression (\ref{SH}) of
the Hamiltonian, the Poisson bracket of a generic higher-order
invariant with ${\Hscr}_S$ can be written as
\begin{equation}\label{poisson2}\eqalign{ 
\{I_J, {\Hscr}_S \} &= {1 \over |S_2|} \{I_J, {\Hscr}_0 \} - 
{{\Hscr}_0 \over |S_2|^2} \{I_J, |S_2| \} \cr 
      &={{\{I_J, {\Hscr}_0 \} - E \{I_J, |S_2| \}} \over {|S_2|}}.\cr} 
\end{equation} 
In the case of a second-order invariant in the standard form of
(\ref{JSI}), this reduces to (\ref{poisson1}). On the other hand, let
us introduce the {\it null} Hamiltonian
\begin{equation}\label{null1} 
{\Hscr}_{\Nscr} = ({\Hscr}_S - E) |S_2| =  
                   {\Hscr}_0 - E |S_2|. 
\end{equation} 
The Poisson bracket of the invariant with ${\Hscr}_{\Nscr}$ is 
\begin{equation}\label{poisson3} 
\{I_J, {\Hscr}_{\Nscr} \} = \{I_J, {\Hscr}_0 \} - 
E \{I_J, |S_2| \}. 
\end{equation} 
Since $|S_2|$ is a non-null positive function everywhere,
eqs.(\ref{poisson2}) and (\ref{poisson3}) are equivalent for what
concernes the conservation of $I_J$. The conceptual difference relies
on the fact that, whereas eq.(\ref{poisson3}) expresses the
conservation of $I_J$ in the dynamics provided by ${\Hscr}_{\Nscr}$ at
{\it zero ``energy"}, eq.(\ref{poisson2}) expresses the conservation
of $I_J$ in the dynamics provided by ${\Hscr}_S$ at {\it arbitrary}
energy. Note that the true energy $E$ enters into ${\Hscr}_{\Nscr}$ as
an arbitrary parameter.
 
In the light of this argument, the procedure to look for new
integrable systems is the following: regard the null Hamiltonian,
written in the form
\begin{equation} 
{\Hscr}_{\Nscr} = {1 \over 2} (p_X^2 + p_Y^2) - \tilde G(X,Y), 
\end{equation} 
where 
\begin{equation}\label{pot} 
\tilde G(X,Y) = A(X) + B(Y) + E |S_2(X,Y)|, 
\end{equation} 
as the system in which to find a new conserved quantity at zero
energy.  If the search of this new conserved quantity is successful,
then {\it it is a strongly conserved quantity for the original
  Hamiltonian ${\Hscr}_S$, which is therefore integrable at arbitrary
  energy}. Going backwards with respect to the previous approach, we
take the potential of eq.(\ref{pot}) and use it to solve the system
(\ref{Kahler}). Once we have the solution, we can write the
integrability condition (\ref{intcon}) in the form
\begin{equation}\label{intcon2} 
\Im \{R_{,ww} \ G +  
3 \ R_{,w} \ G_{,w} + 2 \ R \ G_{,w w} \} = 0. 
\end{equation}  
Note that in the remaining part of this section we suppress the
``tilde" over $\tilde G$ and $\tilde R$ to simplify the notation. The
solution of this equation provides the explicit form of the functions
$A(w + \bar{w})$ and $B(w - \bar{w})$ that allow for the integrability
of the system.  Finally, the integration of the equations for the
trace (\ref{trace}) completes the solution.
 
In the development of this approach, it can be useful to come back to
the separating real variables. Equations (\ref{Kahler}) involving the
K\"ahler potential therefore become:
\begin{equation}\label{KahlerXY}\eqalign{ 
G (X,Y) &= {1 \over 4} \bigl( {\Kscr}_{, XX} + {\Kscr}_{, YY} \bigr),  
\cr 
R (X,Y) &= {1 \over 2} \bigl( {\Kscr}_{, YY} - {\Kscr}_{, XX} \bigr) + 
                          i \ {\Kscr}_{, XY}, \cr} 
\end{equation} 
and the integrability condition (\ref{intcon2}) can be written in the
form
\begin{equation}\label{intcon2XY} 
\eqalign{&\Bigl[ (\Im \{ R \})_{,XX} - (\Im \{ R \})_{,YY} 
            -2 (\Re \{ R \})_{,XY } \Bigr] G +\cr  
       3 &\Bigl[ 
\bigl( (\Im \{ R \})_{,X} - (\Re \{ R \})_{,Y} \bigr) G_{,X} -  
\bigl( (\Re \{ R \})_{,X} + (\Im \{ R \})_{,Y} \bigr) G_{,Y}  
\Bigr] +\cr   
       2 &\Bigl[ 
\Im \{ R \} \bigl( G_{,XX} - G_{,YY} \bigr)  
- 2 \ \Re \{ R \}  G_{,XY} \Bigr] = 0.\cr} 
\end{equation} 
The system (\ref{trace}) finally becomes: 
\begin{equation}\label{traceXY}\eqalign{ 
K_{, X} &= - {4 \over 3} \bigl(  
               (\Re \{ R \})_{,X} + (\Im \{ R \})_{,Y}) \bigr) G  
           - {8 \over 3} \bigl( 
                \Re \{ R \} G_{,X}+ \Im \{ R \} G_{,Y} 
                         \bigr), \cr 
K_{, Y} &= - {4 \over 3} \bigl(  
               (\Im \{ R \})_{,X} - (\Re \{ R \})_{,Y} \bigr) G  
           - {8 \over 3}  \bigl( 
                \Im \{ R \} G_{,X} - \Re \{ R \} G_{,Y} 
                \bigr). \cr } 
\end{equation} 
 
To solve system (\ref{Kahler}), we need only consider the contribution
to the potential (\ref{pot}) coming from the conformal factor:
\begin{equation}\label{hatpot1} 
G_{C} = E \ |S_2(X,Y)|. 
\end{equation}  
Recalling the relation (\ref{dct}) between the conformal
transformation and the $S_2$ function in the quadratic case and the
definition of the inverse transformation (\ref{ict}), the above source
term can also be written as
\begin{equation}\label{hatpot2} 
G_{C} = E \ \sqrt{ S_2(w) \bar S_2 (\bar w)} = E F' (w) \bar F'  
(\bar w). 
\end{equation}  
The solutions of system (\ref{Kahler}) for this term are: 
\begin{equation}\label{hatKahler} 
\eqalign{{\Kscr_{C}}&= 
E \left( F(w) \bar F(\bar w) + 2 \ \Re \{\Lambda (w) \} \right) , \cr 
R_{C} &= - 2 \   
        E \left( F''(w) \bar F(\bar w) + \Lambda'' (w) \right) \cr} 
\end{equation} 
so that the complete solution is therefore 
\begin{equation}\label{Kahlersol} 
\eqalign{\Kscr&=  
4 \int \Bigl[\int A dX \Bigr] dX +  
4 \int \Bigl[\int B dY \Bigr] dY +  
E \left( F(w) \bar F(\bar w) + 2 \Re \{\Lambda 
(w) \} \right) , \cr  
R &= 2 \left(B(Y) - A(X) \right) - 2 \   
     E \left( F''(w) \bar F(\bar w) + \Lambda'' (w) \right). \cr} 
\end{equation} 
This solution is subject to the integrability condition
(\ref{intcon2}).  In analogy with the general approach, substituting
in it the solution (\ref{Kahlersol}) we get a second degree polynomial
in $E$ of the form (\ref{poli}). This equation must be satisfied for
every value of the arbitrary parameter $E$, so that each coefficient
must separately vanish. The three coefficients are in this case:
\begin{equation} 
\eqalign{ 
A_2&= -4 i \ \Im \Bigl\{ (F'''' \bar F + \Lambda'''') F' \bar F'  +  
                        5 F''' F'' \bar F' \bar F +  
                        3 \Lambda''' F'' \bar F' +  
                        2 \Lambda''  F''' \bar F' \Bigr\}, \cr  
A_1&= 4i \ \Im \Bigl\{ 3 (B_{,w} - A_{,w}) F'' \bar F'  
      - 3 (A_{,w} + B_{,w}) \bigl( F''' \bar F + \Lambda''' \bigr)  
     +        2(B - A) F''' \bar F' \cr  
   &\quad - 2 (A_{,ww} + B_{,ww}) \bigl( F'' \bar F + \Lambda'' \bigr) 
     -         (A + B) F'''' \bar F \Bigr\}, \cr 
A_0&= 2i \ \Im \Bigl\{6(B_{,w} - A_{,w})(B_{,w} + A_{,w}) + 4 (B - A) 
(A_{,ww} + B_{,ww}) \Bigr\}.\cr} 
\end{equation} 
From the properties of the functions $A$ and $B$, $A_0$ is
identically zero.  Limiting ourselves to the simplest case $\Lambda =
0$, the condition that $A_2$ vanishes turns out to be
\begin{equation} 
\Im \{ (F'''' F' + 5 F''' F'') \bar F  \bar F' \} = 0, 
\end{equation} 
so that it is necessary and sufficient that 
\begin{equation} 
F'''' F' + 5 F''' F'' = 6aF F', \quad a \in \mathbb{R}. 
\end{equation} 
The general solution is  
\begin{equation}\label{Fsolution} 
(F')^4 = a F^4 + \gamma F^2 + \delta F + \epsilon, 
\end{equation} 
where $\beta, \; \gamma, \; \delta, \; \epsilon$ are complex
constants. We see that solution (\ref{Fsolution}) is a subset of the
general result (\ref{S4}) so that we can state that the simplified
approach presented here consists in choosing a function in the set
\begin{equation} 
S_2(z) = \sqrt{a z^4 + \gamma z^2 + \delta z + \epsilon}. 
\end{equation} 
The equation $A_1=0$ then picks the functions $A$ and $B$ that are
compatible with strong integrability. The practical advantage with
respect to the application of the general technique of section 4 is
that equation $A_0=0$ is automatically satisfied.

 
\subsection{Class $ i z^2$ systems} 
 
To illustrate the simplified approach, in this subsection we present a
class of simple systems which are strongly integrable with a quartic
invariant and, at zero energy, are separable with a quadratic
invariant.
 
Consider then the function  
\begin{equation}\label{iz2} 
S_2(z) = i z^2.  
\end{equation} 
In the solution (\ref{Fsolution}) it corresponds to the choice 
\begin{equation} 
a = -1,  \; 
\gamma = \delta = \epsilon = 0. 
\end{equation} 
This is the simplest polynomial form which does not satisfy the
constraint (\ref{S2}) and therefore gives a potential which is not
automatically integrable at arbitrary energy with a quadratic second
invariant. The corresponding conformal transformation is given by
\begin{equation} 
w = H(z) = {1-i \over {\sqrt2}} \ln z.  
\end{equation} 
Using polar coordinates, so that  
\begin{equation}\eqalign{ 
      z &= x + i y = r \exp({i \theta}) \ , \cr 
      r &= \sqrt{x^2 + y^2}, \ \ \theta = \arctan(y/x), \cr} 
\end{equation} 
in terms of real variables, we get
\begin{equation}\label{eq:XY}\eqalign{ 
      X &= {1 \over {\sqrt2}} (\theta + \ln r) \ ,\cr 
      Y &= {1 \over {\sqrt2}} (\theta - \ln r) \ .\cr 
}\end{equation} 
We point out, in passing, that that in \cite{geom}, in equations  
(102--104) corresponding to eq.\ (\ref{eq:XY}) above, there is 
a misprint and the coordinate $r$ must be substituted with its  
natural logarithm. It follows that the potential given by 
\begin{equation}\label{iz2G} 
      V(\theta, r) = - {{A(\theta + \ln r) + B(\theta - \ln r)} \over  
r^2} \ , 
\end{equation} 
is integrable at zero energy for arbitrary functions $A$ and $B$. The
common factor $ r^{-2} $ is due to the fact that, from the choice of
eq.  (\ref{iz2}), we have
\begin{equation} 
|S(X,Y)| = r^2 = {\rm e}^{\sqrt2 (X - Y)}. 
\end{equation} 
The solutions (\ref{Kahlersol}) are  
\begin{equation} 
\Kscr = 4 \int \Bigl[\int A dX \Bigr] dX +  
        4 \int \Bigl[\int B dY \Bigr] dY + 
        E \ {\rm e}^{\sqrt2 (X - Y)}. 
\end{equation} 
and 
\begin{equation} 
R = 2 \bigl((B(Y) - A(X) \bigr) - 2 \ i \ E \ {\rm e}^{\sqrt2 (X -  
Y)}. 
\end{equation} 
The integrability condition (\ref{intcon2XY}) becomes 
\begin{equation} 
4 \ E \ {\rm e}^{\sqrt2 (X - Y)}  
\bigl[ A'' - B'' + 3 \sqrt2 (A' + B') + 4 (A - B) \bigr] = 0. 
\end{equation} 
This equation splits into 
\begin{equation}\eqalign{A'' + 3 \sqrt{2} A' + 4 A &=0, \cr 
            B'' - 3 \sqrt{2} B' + 4 B &=0, \cr} 
\end{equation} 
with solutions: 
\begin{equation}\eqalign{A(X) &= a_1 {\rm e}^{- \sqrt2 X} + 
                    a_2 {\rm e}^{- 2 \sqrt2 X}, \cr 
            B(Y) &= b_1 {\rm e}^{ \sqrt2 Y} + 
                    b_2 {\rm e}^{ 2 \sqrt2 Y}. \cr} 
\end{equation} 
The potential  
\begin{equation} 
V(r, \theta) = - 
{{a_1 {\rm e}^{- \theta} + b_1 {\rm e}^{ \theta}} \over {r^3}} -  
{{a_2 {\rm e}^{- 2 \theta} + b_2 {\rm e}^{2 \theta}} \over {r^4}} 
\end{equation} 
ie therefore integrable. As a simple concrete example, let us take 
\begin{equation} 
a_1 = b_1 = 0, \; a_2 = b_2 = - {1 \over 2}, 
\end{equation} 
so that the potential is 
\begin{equation} 
V = {{{\rm e}^{2 \theta} + {\rm e}^{- 2 \theta}} \over {2 r^4}}, 
\end{equation} 
which is a known result mentioned in the review by Hietarinta. The
second invariant has the form
\begin{equation} 
I = p_{\theta}^4 + {\rm ch} 2 \theta p_r^2 + 2  
{{{\rm sh} 2 \theta} \over {r}} p_r p_{\theta} + 3  
{{{\rm ch} 2 \theta} \over {r^2}} p_{\theta}^2  
+ 2 {{2-{\rm sh} 2 \theta} \over {r^4}}. 
\end{equation}

 
\section{Classification of Hamiltonians admitting a quartic invariant  
at 
arbitrary energy} 
 
In the tables included in this section we present all the solutions we
have found of systems admitting an independent invariant quartic in
the momenta. Some of them are new. For all cases we present the
potential $V$ (up to linear transformations of the coordinates and a
linear rescaling of the potential itself) as well as the quartic
invariant $I$, thereby making it possible to directly compare our
results with Hietarinta's classification of 1987 \cite{hiet}. For the
cases which are superintegrable we also indicate, using the notation
of \cite{hiet}, which of the real quadratic cases (1), (2), (4) or (7)
a given system belongs to.
 
A peculiar property of our solutions is that all but one have a
conformal factor generated by function $S$ of even degree in $z$, with
real coefficients. As a consequence of this, the solutions obtained
can be expressed in terms of the coordinate systems of separable
potentials. In fact, since from (\ref{transf}) and (\ref{ict}), the
transformation is
\begin{equation}\label{transf4} 
H(z) = F^{-1} (w(z)) = \int dz \ [S_4(z)]^{-1/4}, 
\end{equation} 
with $S$ of the form 
\begin{equation} 
S_4(z) = \pm (a z^2 + \beta z + \gamma), \quad 
a \in \mathbb{R}, \quad \beta, \gamma \in \mathbb{C},  
\end{equation} 
we are actually taken back to the coordinate transformations to  
separable variables given by an $S_2$ function satisfying the  
condition (\ref{S2}), for which the transformation is of the form  
\cite{geom} 
\begin{equation}\label{transf2} 
H(z)= \int dz \ [S_2(z)]^{-1/2}. 
\end{equation} 
As an exception to this situation, we have also found a solution with  
$S_4$ of third degree in $z$, by the method of the {\it coupling  
constant metamorphosis} (see \cite{hiet}) which is described in  
Section 6.2 below.

\subsection{Systems for which $S(z)$ is of even degree} 
 
The simplest case is that with $S_4$ of degree zero. This case
corresponds simply to rotations and translations of the Cartesian
coordinates. In table 1 we list the new solutions obtained and, for
comparison, the quartic integrable potential of \cite{gdr}. In table 2
the already known solutions of this class are listed together with the
proper references. The first sequence of items correspond to the
simple solution sketched at the end of section 4. All solutions are
generated by $S_4 = \pm 1$: actually, there is nothing fundamental in
the choice of sign which has been made only for sake of comparison
with already known results.
 
We then have some solutions with $S_4$ of degree two and four in $z$.
In table 3, two new interesting classes of integrable systems are
presented with $S_4$ of degree two. They are more conveniently treated
using the parabolic coordinates provided by the conformal
transformation generated by $S_4 = 16 z^2$. In table 5, a class of
integrable systems generated by $S_4 = -z^4$ is presented,
generalizing the separability in polar coordinates. Here, the choice
of sign is dictated by the transformation used in the simplified
approach of section 5.3 above, but to the same system one can also
arrive with the general method using the ``proper'' transformation
given by $S_4 = z^4$.
 
\subsection{A system for which $S(z)$ is of degree three} 
 
In order to describe how we get this new solution, we briefly recall  
the method based on the coupling constant metamorphosis \cite{hiet}.   
We introduce a null Hamiltonian which, in analogy with  
eq.(\ref{null1}), is defined as 
\begin{equation}\label{null2} 
{\Hscr}_{\Nscr} = ({\Hscr} - E) |S_4|^2. 
\end{equation} 
The second case of table 3 has a null Hamiltonian which, if expressed  
in the new coordinates, has the form 
\begin{equation} 
{\Hscr}_{\Nscr} = {1 \over 2} (p_X^2 + p_Y^2) - A(X) - B(Y)  
                  + 4 C (X^2 + Y^2)^3 - 4 E (X^2 + Y^2). 
\end{equation} 
Then, if the term $(X^2 + Y^2)^3$ can be identified with the conformal
factor associated to a new conformal transformation, we may interpret
$C$ as a {\it new} energy and $E$ as a {\it new} coupling constant. We
get a new potential which is strongly integrable with an invariant of
the same nature as before. This is the process known as coupling
constant metamorphosis.
 
In the present instance, we have that the transformation must be  
still of the form (\ref{transf4}) or, equivalently,  
\begin{equation}\label{ccm1} 
F'(w)= S_4^{1/4}, 
\end{equation} 
where, for simplicity, we still denote the new variable as $w = X + i
Y$, so that $|w|^2 = X^2 + Y^2$. Since the conformal factor is assumed
to be $(X^2 + Y^2)^3$, we get
\begin{equation}\label{ccm2} 
|F'|^2 \sim |w|^6. 
\end{equation} 
Therefore, $F(w)$ must be of the order $w^4$, so that  
\begin{equation}\label{ccm3} 
w = H(z) \sim z^{1/4}. 
\end{equation} 
From (\ref{transf4}) we get,  
\begin{equation}\label{ccm4} 
H'(z) \sim z^{-3/4}, 
\end{equation} 
so that, finally, 
\begin{equation}\label{ccm5} 
S_4(z) \sim z^{3}. 
\end{equation} 
We end up with a solution of third degree for $S$ which had escaped a
previous direct analysis. The new integrable Hamiltonian is
\begin{equation} 
\eqalign{{\Hscr} = &{1 \over 2} (p_r^2 + {{p_{\theta}^2} \over  
                     {r^2}}) + {{\mu} \over {r}} +\cr 
                         & {{a} \over {r^2 (1 + \cos {\theta \over 2})}}  
                         + {{b} \over {r^2 (1 - \cos {\theta \over 2})}}  
                         + {{c} \over {r^3 (1 + \cos {\theta \over 2})^3}}  
                         + {{d} \over {r^3 (1 - \cos {\theta \over 2})^3}} = {\tilde E},  
\cr} 
\end{equation} 
where ${\tilde E} = C$ is the new energy and $\mu = - E$ is the new
coupling constant. Table 4 reports the form of the coordinate
transformation and the expression of the second invariant.

 
  
 \begin{table}[tbp] 
 \begin{tabular}{l} 
 \Hline\Hline 
 $S(z) = 1$ \\ 
         $\Kscr = 2E x^2 - {4 \over 3} x^4 - 2 b x y^2  
         - {1 \over 3} y^4 + 4c \ln |y|$ \\ 
         $V = 4 x^2 + y^2 + b x + c y^{-2} $       \\ 
         $I = p_x^4 + 2 p_x^2 p_y^2 +  
     4 V p_x^2 + 4 b y p_x p_y + 16 x^2 p_y^2 + 2 (2 x^2 + y^2)(b + 4x)^2 + 32 c x^2 
     y^{-2}$ \\ 
  Superintegrable, case (4) and (7) \\ 
 \hline\hline 
 $S(z) = 1$ \\ 
         $\Kscr = 2E x^2 - 9 y^{4/3} (b x + c) - {3 \over 2} a y^4 -  
          {9 \over 10} d y^{8/3} - {8 \over 3} a x^4$ \\ 
         $V = {1 \over 2} a (16 x^2 + 9 y^2) + (b x + c) y^{-2/3} + d 
        y^{2/3} $       \\ 
         $I = p_x^4 + 2 p_x^2 p_y^2 +  
     4 V p_x^2 + 12 b y^{1/3} p_x p_y + 32 a x^2 p_y^2 + 64 a x^2 V $ \\ 
  $\quad  + 16 b d x - 256 a^2 x^4 + 18 b^2 y^{2/3} + 144 a b x y^{4/3}$ \\  
 \hline\hline 
 $S(z) = -1$ \\ 
         $\Kscr = 2E x^2 - 2x^4y^2-{1 \over 3}a x^4y -  
   {{a^2} \over {12}} \left( {{x^4} \over 18} + y^4 \right) - {16 \over 15} y^6 - 
   {8 \over 15} a y^5 + 4d \ln |x|$ \\ 
         $\displaystyle V = x^4 + 6 x^2 y^2 + 8 y^4 +a \left({8 \over 3} y^3 + x^2 y \right) 
     + {{a^2} \over {72}} \left(x^2 + 16 y^2 \right) + {{d} \over {x^2}} $, \\ 
         $\displaystyle I = p_x^4 +  
     4 p_x^2 \left(x^4 + 6 x^2 y^2 + a x^2 y + {1 \over 72} a^2 x^2 + {d \over {x^2}} 
             \right) - \left({4 \over 3} a x^3 + 16 x^3 y \right) p_x p_y $ \\ 
  $\quad + 4 x^4 p_y^2 + 4 \bigl(2 d  x^2 + x^8 + d^2 x^{-4} 
     + {2 \over 3} a d y + 4 d y^2 \bigr) +   $ \\ 
  $\displaystyle \quad {{x^6} \over 9} \left(144 y^2 + 24 a y - a^2 \right) + 
         {{x^4} \over {1296}} {\left(144 y^2 + 24 a y - a^2 \right)}^2 $ \\  
 \hline\hline 
 $S(z) = -1$ \\ 
         $\Kscr = 2E x^2 -2x^4y^2 - {16 \over 15} y^6 - 
   {1 \over 3} b x^4 - {4 \over 3} b y^4  + 4d \ln |x| - {c \over 5} y^{-4}  
    + 4m \ln |y|$ \\ 
         $\displaystyle V = x^4 + 6 x^2 y^2 + 8 y^4 + b (x^2 + 4 y^2) + {{d} \over {x^2}} +  
      {{m} \over {y^2}} + {{c} \over {x^6}}$, (Grammaticos et al. 1984) \\ 
         $\displaystyle I = p_x^4 +  
     4 p_x^2 \left(x^4 + 6 x^2 y^2 + b x^2 + {{d} \over {x^2}} + {{c} \over {x^6}}  
             \right) - 16 x^3 y p_x p_y + 4 x^4 p_y^2 $ \\ 
  $\quad + 4 \bigl[12 c y^2 x^{-4} + 2d x^2 + x^8 + 2b x^6 + b^2 x^4   
     + 2c x^{-2} + 2 cd x^{-8} + c^2 x^{-12} + (d^2 + 2bc) x^{-4} $ \\ 
  $\quad + 4 x^6 y^2 + 2m x^4 y^{-2} + 4b x^4 y^2 + 4 x^4 y^4 
         + 4d y^2 \bigr] $ \\  
 \Hline\Hline 
 \end{tabular} 
         \caption{\protect\small\protect\parbox[t]{12cm}{ 
           Systems for which $S(z)$ is of zero degree.}} 
         \protect\label{tab:Szero} 
 \end{table} 
  
 \newpage 
  
 \begin{table}[tbp] 
 \begin{tabular}{l} 
 \Hline\Hline 
 $S(z) = 1$ \\ 
         $\Kscr = E (x^2 + y^2)-f_1(x)-f_2(y)-f_3(x-y)-f_4(x+y)$ \\ 
         $V = v_1(x)+v_2(y)+v_3(x-y)+v_4(x+y)$, \\ 
         \quad $v_i(\xi) = \frac14 f_i''(\xi)$\, (for $i=1,2$), $v_i(\xi) = \frac12
 f_i''(\xi)$\, (for $i=3,4$),\, (Hietarinta, 1987) \\
         $I = p_x^2 p_y^2 + 2 v_2 p_x^2 + 2 (v_4 - v_3) p_x p_y + 2 v_1 p_y^2 + h(x,y)$ \\ 
  $v_1=0, \; v_2={\rm e}^y, \; v_3={\rm e}^{x-y}, \; v_4=0, \;  
   h=2{\rm e}^x+{\rm e}^{2(x-y)}.$ \\ 
 \hline 
  $v_1={\rm e}^{-x}, \; v_2={\rm e}^y, \; v_3={\rm e}^{x-y}, \; v_4=0, \;  
   h=2{\rm e}^x+{\rm e}^{2(x-y)}+2{\rm e}^{-y}+4{\rm e}^{y-x}.$ \\ 
 \hline 
  $v_1=0, \; v_2={\rm e}^y, \; v_3={\rm e}^{x-y}, \; v_4={\rm e}^{-x-y}, \;  
   h=2{\rm e}^x+{\rm e}^{2(x-y)}+2{\rm e}^{-x}+{\rm e}^{-2(x+y)}-2{\rm e}^{-2y}.$ \\ 
 \hline 
  $v_1=0, \; v_2={\rm e}^y, \; v_3={\rm e}^{x-y}, \; v_4={\rm e}^{-(x+y)/2}, \;  
   h=2{\rm e}^x+{\rm e}^{2(x-y)}+{\rm e}^{-(x+y)}-2{\rm e}^{(x-3y)/2}.$ \\ 
 \hline 
  $v_1={\rm e}^{-2x}, \; v_2={\rm e}^y, \; v_3={\rm e}^{x-y}, \; v_4=0, \;  
   h=2{\rm e}^x+{\rm e}^{2(x-y)}+4{\rm e}^{-2x+y}.$ \\ 
 \hline 
  $\displaystyle v_1={{a} \over {x^2}}, \; v_2={{a} \over {y^2}}, \;  
   v_3={{b} \over {(x-y)^2}}, \; v_4={{b} \over {(x+y)^2}},\; 
   h={{4 a^2} \over 
 {x^2 y^2}} + {{16 b^2 x^4} \over {(x^2-y^2)^4}} -  
     {{16 b^2 x^2} \over {(x^2-y^2)^3}} + 
     {{16 ab} \over {(x^2-y^2)^2}} .$ \\ 
 \hline\hline 
 $S(z) = 1$ \\ 
         $\Kscr = 2E x^2 - 9y^{4/3} (c x^2 + d + {m \over 10} 
   y^{4/3}) - a ({4 \over 3} x^4 + 3 y^4) + 4n \ln |x|$ \\ 
         $\displaystyle V = {9 \over 2} c y^{4/3} + (c x^2 + d) y^{-2/3} + m y^{2/3}  
     + a (9 y^2 + 4 x^2) + {n \over {x^2}}$, (Grammaticos et al. 1984)        \\ 
         $\displaystyle I = p_x^4 + 2 p_x^2 p_y^2 +  
     4 p_x^2 \left((c x^2 + d) y^{-2/3}+ m y^{2/3}  
     + a (9 y^2 + 4 x^2) + {n \over {x^2}}\right) + 24 c x y^{1/3} p_x p_y $ \\ 
  $\displaystyle \quad + 4 p_y^2 (4 a x^2 + n x^{-2}) + 16 c m x^2 + 32 a d x^2 y^{-2/3}  
     + 8 {{d n} \over {x^2 y^{2/3}}} $ \\ 
  $\displaystyle \quad + 8 {c n \over {y^{2/3}}} + (32 a m + 72 c^2)  
       x^2 y^{2/3} + 8 n m x^{-2} y^{2/3} + 72 a n x^{-2} y^2    $ \\ 
  $\quad + 4 n^2 x^{-4} + 32 a c x^2 y^{-2/3} (9 y^2 + x^2)  
         + 32 a^2 x^2 (9 y^2 + 2 x^2) $ \\  
 \hline\hline 
 $S(z) = -1$ \\ 
         $\Kscr = 2E x^2 - a \left({1 \over 6}x^4 + {8 \over 3} y^4 \right)  
   -d \left({1 \over 3}x^4y + {16 \over 15} y^5 \right) + 4h_1 \ln |x| - 
   {{h_2} \over {5 x^4}}$ \\ 
         $\displaystyle V = {1 \over 2} a (x^2 + 16 y^2) + d \left({16 \over 3} y^3 + x^2 y  
         \right) 
     + {{h_1} \over {x^2}} + {{h_2} \over {x^6}}$, (Grammaticos et al. 1984) \\ 
         $\displaystyle I = p_x^4 +  
     4 p_x^2 \left({1 \over 2} a x^2 + d x^2 y  
     + {{h_1} \over {x^2}} + {{h_2} \over {x^6}} \right)  
     - {4 \over 3} d x^3 p_x p_y $ \\ 
  $\displaystyle \quad + 4 \Bigl[2 d h_2 y x^{-4}  
     + 2 h_1 h_2 x^{-8} + h_2^2 x^{-12} + (h_1^2 + a h_2) x^{-4} 
     + {1 \over 4} a^2 x^4 - {1 \over 18} d^2 x^6   $ \\ 
  $\quad - {1 \over 3} d^2 x^4 y^2 - {1 \over 3} a d x^4 y 
         + {2 \over 3} d h_1 y \Bigr] $ \\  
 \Hline\Hline 
 \end{tabular} 
         \caption{\protect\small\protect\parbox[t]{12cm}{ 
           Systems for which $S(z)$ is of zero degree (continued).}} 
         \protect\label{tab:Szero-2} 
 \end{table} 
  
 \newpage 
  
 \begin{table}[tbp] 
 \begin{tabular}{l} 
 \Hline\Hline 
 $S(z) = 16 z^2$ \\ 
         $\Kscr = \frac23 E (2x^2 + y^2)+f(X)+g(Y) + C (X Y)^{\alpha}$ \\ 
  $ X = \sqrt{{r+x}\over 2}, \;\;  
    Y = \sqrt{{r-x}\over 2}, \;\; r = \sqrt{x^2+y^2} = X^2+Y^2.$\\ 
 \hline\hline 
 $ f(X) = - 36 \,\, 2^{-1/3} c X^{4/3} - {{18\,\, 2^{1/3}} \over {5}} a X^{8/3}, \;\; 
   g(Y) = - 36 \,\, 2^{-1/3} d Y^{4/3} - {{18\,\, 2^{1/3}} \over {5}} b Y^{8/3}, \;\; 
    \alpha = 4/3.$\\ 
         $\displaystyle V = {{a (r+x)^{1/3} + b (r-x)^{1/3} + c (r+x)^{-1/3} + d (r-x)^{-1/3}}\over r} 
     + {{k} \over {y^{2/3}}}, \; \; k = - \frac{2^{1/3} C}{36},$\\ 
   
         $I = J^2 - 8 k y^{1/3} \left[ p_x p_{\theta} - {y \over r}  
    (a (r+x)^{1/3} + b (r-x)^{1/3} + c (r+x)^{-1/3} + d (r-x)^{-1/3}) \right]$ \\ 
  $\quad - 8 k\left[a (r-x)^{2/3} + b (r+x)^{2/3} \right],$ \\ 
  $J = p_y p_{\theta} + 
   {{(r+x) [b (r-x)^{1/3} + d (r-x)^{-1/3}] -  
     (r-x) [a (r+x)^{1/3} + c (r+x)^{-1/3}]}\over r} + 2{{k x} \over {y^{2/3}}}$, \\ 
  $p_{\theta} = x p_y - y p_x. $ \\  
 \hline\hline 
  $ f(X) = {1 \over 14} C X^8 + 8a \log X - {1 \over {10}} c X^{-4}, \;\; 
    g(Y) = {1 \over 14} C Y^8 + 8b \log Y - {1 \over {10}} d Y^{-4}, \;\; 
    \alpha = 4.$\\ 
         $\displaystyle V = {1 \over r}  
       \left[{a \over {r+x}} +  
             {b \over {r-x}} +  
             {c \over {(r+x)^3}} +  
             {d \over {(r-x)^3}} \right] 
     + {1 \over 2} k r^2, \; \; k=-\frac{C}{2},$\\
 $I = J^2 - k y^3 \left[ p_x p_{\theta} - {y \over r}  
    \left({a \over {r+x}} +  
             {b \over {r-x}} +  
             {c \over {(r+x)^3}} +  
             {d \over {(r-x)^3}} \right) + {1 \over 4} k x^2 y\right]$ \\ 
  $\quad - k\left[a (r-x)^{2} + b (r+x)^{2} \right],$ \\ 
  $J = p_y p_{\theta} + 
   {1 \over r} \left[ (r+x) [{b \over {r-x}} + {d \over {(r-x)^3}}] -  
                      (r-x) [{a \over {r+x}} + {c \over {(r+x)^3}}] \right] +  
      {1 \over 2} k x y^2 .$   \\  
 \Hline\Hline 
 \end{tabular} 
         \caption{\protect\small\protect\parbox[t]{12cm}{ 
           Systems for which $S(z)$ is of second degree.}} 
         \protect\label{tab:Stwo-2} 
 \end{table} 
  
 \begin{table}[tbp] 
 \begin{tabular}{l} 
 \Hline\Hline 
  $S(z) = 64 z^3$ \\ 
         $\Kscr = \frac27 E(4x^2+3y^2)+f(X)+g(Y) - \frac{4}{3}\mu(X^4+Y^4)$ \\ 
  $  X = {\sqrt2}\,r^{1/4} \cos {\theta \over 4} = r^{1/4}\sqrt{1+\cos
  {\theta \over 2}}, \ \  
     Y = {\sqrt2}\,r^{1/4} \sin {\theta \over 4} = r^{1/4}\sqrt{1-\cos
  {\theta \over 2}}, $ \\ 
    $r = \sqrt{x^2+y^2} = (X^2+Y^2)^2/4, \ \  
     \theta = \arctan(y/x). $\\ 
 \hline\hline 
  $ f(X) = 32 a \ \log X - {8 \over 5} c \ X^{-4},$ \\
  $ g(Y) = 32 b \ \log Y - {8 \over 5} d \ Y^{-4}.$ \\

 $\displaystyle V   = {{\mu} \over {r}} + {{a} \over {r^2 (1 + \cos {\theta \over 2})}}  
                      + {{b} \over {r^2 (1 - \cos {\theta \over 2})}}  
                      + {{c} \over {r^3 (1 + \cos {\theta \over 2})^3}}  
                      + {{d} \over {r^3 (1 - \cos {\theta \over 2})^3}}.$\\ 
 $I = J^2 + \Gamma_{1} \Pi - \Gamma_{2},$\\ 
  
 $J = \Delta + B(Y) - A(X) - {\Hscr} (X^2 - Y^2)^3 + 4 \mu (X^2 - Y^2),$\\ 
  
 $\Delta = {{2} \over {\sqrt{r}}}  
 \left[ \cos {\theta \over 2} (r^2 p_{r}^2 - p_{\theta}^2)  
       -\sin {\theta \over 2} r p_{r} p_{\theta} \right],$\\ 
 $\Pi = {{2} \over {\sqrt{r}}}  
 \left[ \sin {\theta \over 2} (r^2 p_{r}^2 - p_{\theta}^2)  
       +\cos {\theta \over 2} r p_{r} p_{\theta} \right],$\\ 
  
 $\Gamma_{1} = 16 \ {\Hscr} \ X^3 Y^3 ,$\\ 
  
 $\Gamma_{2} = 16 \ {\Hscr} \ [{\Hscr} \ (2 X^6 Y^6 + 3 X^4 Y^4 (X^4 + Y^4)) 
 - 8 \mu X^4 Y^4  - a Y^4 - b X^4].$\\
 $ {\Hscr} $ is the Hamiltonian of the system, i.e.\  
${\Hscr} = {1 \over 2} (p_r^2 + {{p_{\theta}^2} \over {r^2}})+V,$ \\
$p_{\theta} = x p_y - y p_x, $ \\ 
 $r p_r = x p_x + y p_y. $ \\ 
 \Hline\Hline 
 \end{tabular} 
         \caption{\protect\small\protect\parbox[t]{12cm}{ 
           Systems for which $S(z)$ is of third degree.}} 
         \protect\label{tab:Stwo-3} 
 \end{table}

 \begin{table}[tbp] 
 \begin{tabular}{l} 
 \Hline\Hline 
  $S(z) = - z^4$ \\ 
         $\Kscr = E r^2 +f(X)+g(Y)$ \\ 
  $  X = {1 \over {\sqrt2}} (\theta + \ln r) , \ \  
     Y = {1 \over {\sqrt2}} (\theta - \ln r) . $ \\ 
 \hline\hline 
 $ A(X) = a_1 {\rm e}^{- \sqrt2 X} + 
          a_2 {\rm e}^{- 2 \sqrt2 X}, \ \ A = f''/4$ \\ 
 $ B(Y) = b_1 {\rm e}^{ \sqrt2 Y} + 
          b_2 {\rm e}^{ 2 \sqrt2 Y},  \ \ B = g''/4.$\\  
         $\displaystyle V = {{a {\rm e}^{  \theta} + b {\rm e}^{-   \theta}} \over {r^3}} +  
       {{c {\rm e}^{2 \theta} + d {\rm e}^{- 2 \theta}} \over {r^4}},$ \\ 
         $I = p_{\theta}^4 +  
      (c {\rm e}^{2 \theta} + d {\rm e}^{- 2 \theta}) p_r^2 +  
     2\left[a {\rm e}^{\theta} - b {\rm e}^{-\theta} + {1 \over r}  
      (c {\rm e}^{2 \theta} - d {\rm e}^{- 2 \theta}) \right]  
       p_r p_{\theta}$ \\ 
 $\quad + \left[{2 \over r} (a {\rm e}^{\theta} + b {\rm e}^{-\theta}) +  
              {3 \over r^2} (c {\rm e}^{2 \theta} + d {\rm e}^{- 2 \theta}) 
          \right] p_{\theta}^2$ \\ 
 $\quad - {1 \over r^2} (a {\rm e}^{\theta} - b {\rm e}^{-\theta})^2 + 
          {4 \over r^3} (bc {\rm e}^{\theta} + ad {\rm e}^{- \theta}) + 
          {1 \over r^4} (c^2 {\rm e}^{4 \theta} + d^2 {\rm e}^{- 4 \theta} 
           + 6cd).$\\ 
 
 \Hline\Hline 
 \end{tabular} 
         \caption{\protect\small\protect\parbox[t]{12cm}{ 
           Systems for which $S(z)$ is of fourth degree.}} 
         \protect\label{tab:Stwo-4} 
 \end{table} 
  
 
 
\section{Conclusions} 
 
In this work we have explored the set of natural 2--dimensional
Hamiltonian systems admitting a second invariant which is a polynomial
in the momenta of degree four. The approach is based on the Jacobi
geometrization described in \cite{geom}, where the linear and
quadratic cases were also explored, and in \cite{max}, where the cubic
cases were treated. The approach allows a unified treatment of
integrability at fixed and arbitrary energy, even if here we have
limited ourselves to obtaining the list, as complete as possible, of
strongly integrable systems. There are several areas in which this
approach can be still useful. The most natural seems that of looking
for systems admitting higher order invariants of which there exist
only a few known examples.
 
Moreover, other properties of invariants up to degree four are still
worth of analysis. In \cite{pr} we have examined the properties of
weakly integrable systems with quadratic second invariants and found
interesting behaviours both in the integrable regime and in the
generic non-integrable one. One may wonder about the information
higher order weak invariants can give on the dynamics of
non-integrable systems. It is reasonable to believe that increasing
the order can improve the reliability of these functions when
interpreting them as effective invariants approximately conserved in
regular portions of phase-space. It would also be interesting to look
for other examples of quartic invariants which correspond to lower
rank Killing tensors on fixed energy surfaces.
 

\end{document}